\documentclass{he_symp}
\usepackage{psfig,graphicx,epsfig}
\usepackage{color}
\usepackage{amsmath,amssymb,epic,eepic,array}
\begin{document}
\renewcommand{\FirstPageOfPaper }{ 58}\renewcommand{\LastPageOfPaper }{ 63}%%
%% This is he_symp_example.tex
%% LaTeX2e example style file for the 270. Heraeus Seminar on 
%% Neutron Stars, Pulsars and Supernova Remnants, held in Bad Honnef, Jan. 21-25, 2002 
%% Needs the LaTeX2e class file he_symp.cls
%%
% -----------------------------------------------------------------------------
%\documentclass{he_symp}
%\usepackage{psfig}
% -----------------------------------------------------------------------------
%\def\R{~ROSAT}
%\def\RAS{\R all sky survey}
\def\cha{{\sl Chandra}}
% -----------------------------------------------------------------------------
%\begin{document}

\title{Chandra Observations of Neutron Stars~--- An Overview}
\author{M.C.Weisskopf }  
\institute{NASA Marshall Space Flight Center, MSFC/SD50, Huntsville, AL 35812, USA}
\maketitle

\begin{abstract}
We present a brief review of \cha\ observations of neutron stars, with a concentration on neutron stars in supernova remnants. The early Chandra results clearly demonstrate how critical the angular resolution has been in order to separate the neutron star emission from the surrounding nebulosity. 
\end{abstract}

\section{Introduction}

\begin{figure}
\centerline{\psfig{file=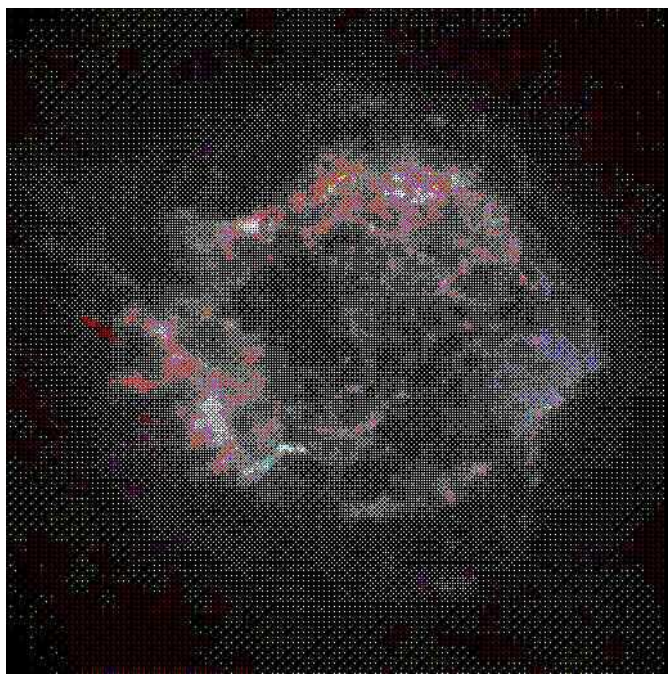,width=8.8cm,clip=} }
\caption{\cha\ ACIS spectrometric image of the Cas-A SNR. 
The red, green, and blue regions in this 6-arcmin-square image show the intensity of low-, medium-, and high-energy X rays, respectively. 
The red material on the east outer edge is enriched in iron, whereas the bright greenish white region on the southeast is enriched in silicon and sulfur. 
In the blue region to the west, low- and medium-energy X rays have been filtered out by a cloud of dust and gas. 
Courtesy NASA/CXC/SAO/Rutgers/Hughes et al. (2000).
\label{f:casa}}
\end{figure}

Beginning with the now-famous ``first-light'' observation of the supernova remnant (SNR) Cassiopeia A and discovery of its central object (Fig.~\ref{f:casa}), it was evident that the {\sl Chandra X-ray Observatory}'s superb angular resolution would play an important role in advancing our understanding of neutron stars. 
Now, more than two years into the mission, this promise has been clearly fulfilled, although the full significance of even the currently completed observations is not yet fully realized. 
Here, I provide a limited (and biased) overview of several \cha\ observations of neutron stars. 
In nearly all cases, I summarize work by others, as attributed in the text and figure captions. 
In addition, I acknowledge the dedication of the entire \cha\ science team~--- at the the Smithsonian Astrophysical Observatory, the Marshall Space Flight Center, and the institutions of the Principal Investigators and their colleagues~--- that has made the \cha\ mission a such a success.
  
\section{Neutron Stars in SNR}

\subsection{Crab Pulsar}

\begin{figure}
\centerline{\psfig{file=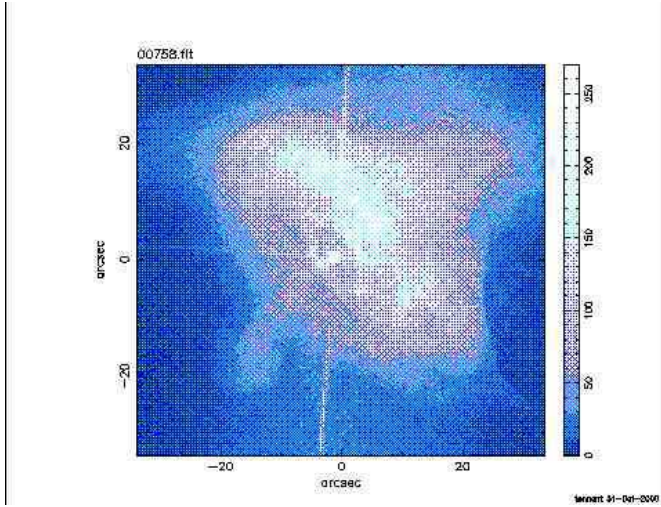,angle=90,width=8.8cm,clip=} }
\caption{
\cha\ LETG--HRC-S zeroth-order image of the Crab Nebula and Pulsar. 
The nearly vertical line is the dispersed spectrum of the pulsar.
The nearly horizontal line is the cross-dispersed spectrum produced by the LETG fine support bars. 
\label{f:crab}}
\end{figure}

\begin{figure}
\centerline{\psfig{file=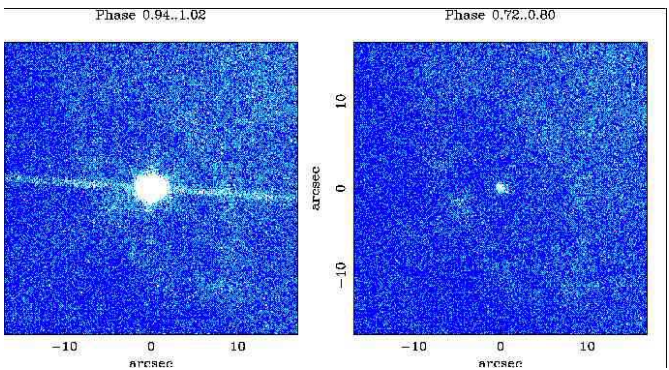,width=8.8cm,clip=} }
\caption{
LETG--HRC-S zeroth-order images of the Crab Pulsar and inner nebula (ObsID 758) at pulse maximum (left) and at pulse minimum (right). 
The X-ray pulsar is clearly discernible at all pulse phases.
\label{f:crab_on-off}}
\end{figure}

\begin{figure}
\centerline{\psfig{file=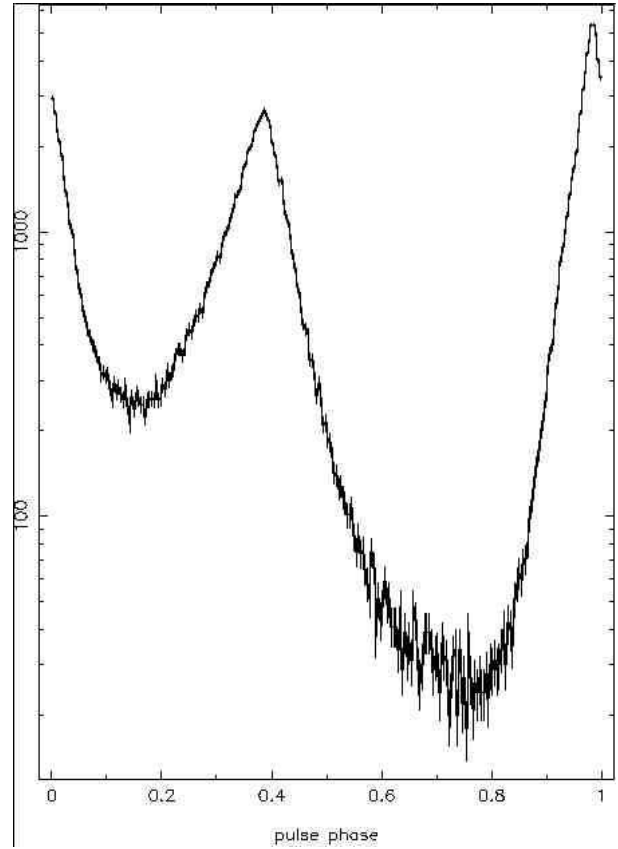,width=8cm,clip=} }
\caption{
The light curve of the Crab pulsations from Tennant et al. (2001).
\label{f:lightcurve}}
\end{figure}

A critically important example of \cha's ability to provide high-contrast images of low-surface-brightness features is the now-classic image (Fig.~\ref{f:crab}) of the Crab Nebula and Pulsar (Weisskopf et al. 2000), which shows the complex spatial structure of the inner nebula.
This image revealed an X-ray inner ring between the pulsar and the X-ray torus.
In Tennant et al.\ (2001), our phase-resolved analysis of an LETG--HRC-S zeroth-order image discovered significant X-ray emission from the pulsar in its ``off'' phase (Fig.~\ref{f:crab_on-off}).  
Owing to the high surface brightness and complex structure of the Crab's inner (plerionic) nebula, \cha's high angular resolution was {\em essential\,} to these discoveries.
We used the measurement of the pulsar's flux at pulse minimum to establish an upper limit to thermal emission from the surface of the neutron star. 
At that time we argued that, for a number of reasons, the observed flux was likely to be predominantly non-thermal. 
First, the detected visible flux at pulse minimum (Golden, Shearer, \& Beskin 2000) demonstrates the presence of unresolved non-thermal emission at pulse minimum. 
Second, the similarity of the relative amplitudes of the X-ray and visible pulsed emission served as another reason~--- in particular, the ratios of flux at pulse minimum to that at pulse maximum are similar in both bands. 
The absence of a pronounced floor in the X-ray pulse profile shown in Fig.~\ref{f:lightcurve} is another, qualitative consideration. 

\begin{figure}
\centerline{\psfig{file=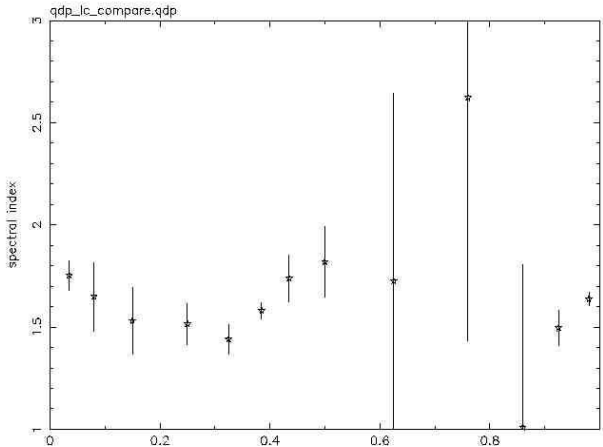,width=8.8cm,clip=} }
\caption{
Spectral index of the Crab Pulsar versus pulse phase.
\label{f:alpha_phase}}
\end{figure}

Subsequently, we (Weisskopf et al.\ 2002, in preparation) analyzed the dispersed data (ObsID 759), studying the spectral evolution as a function of pulse phase. 
Our analysis shows a variation of power-law spectral index with pulse phase (Fig.~\ref{f:alpha_phase}) that is in semi-quantitative agreement with previous measurements (e.g., Pravdo, Angelini, \& Harding 1997; Massaro et al. 2000) at various different energies. The Chandra results extend the phase coverage through pulse minimum, in that prior observations had to treat the data at pulse minimum as background -- a necessary, but incorrect, assumption. 
These analysis show that there is no reason to invoke a thermal component, and result in an independent upper limit similar to that based only on the imaging data. 
Unfortunately, neither limit challenges current theories of neutron star cooling.
Nevertheless, analysis of these data suggest a new experiment, which we are proposing for Chandra Cycle-4, that can efficiently accomplish this objective. 

\begin{figure}
\centerline{\psfig{file=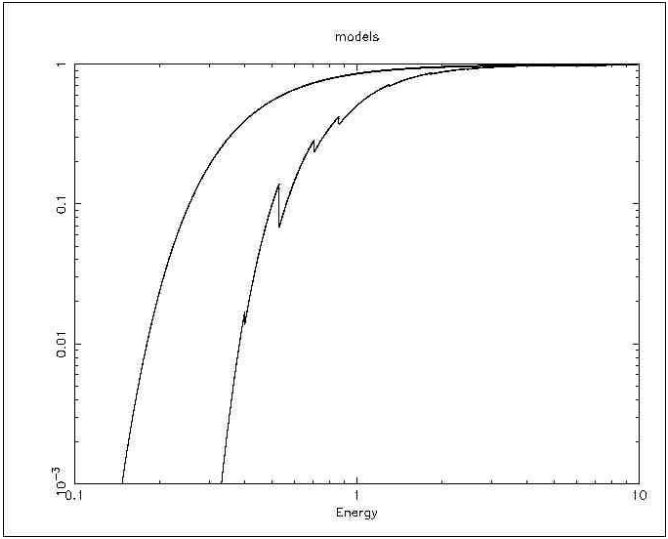,angle=90,width=8.8cm,clip=} }
\caption{
Spectral dependence of dust scattering (smooth curve) and of interstellar photo-electric absorption. 
\label{f:models}}
\end{figure}

As part of the spectral analysis, our colleague Frits Paerels emphasized that that scattering of point-source radiation by interstellar dust grains produces a wavelength-dependent (both in intensity and width) scattering halo (Mauche \& Gorenstein 1986, and references therein) and that this scattering also requires an aperture correction to the effective-area curve. 
Fig.~\ref{f:models} compares the energy dependence of interstellar absorption with that of dust scattering. 
The characteristic scattering angle is of order 10~arcmin and scales approximately as $(1\, {\rm keV)/E}$.
Thus, we effectively measure only the unscattered flux from the point source in our narrow aperture, so that dust scattering produces an energy-dependent modulation of the point-source continuum.
Failure to account for dust scattering in the spectral fitting {\em overestimates} the hydrogen column, in this case by about 20\%. 

\subsection{Vela Pulsar}

\begin{figure}
\centerline{\psfig{file=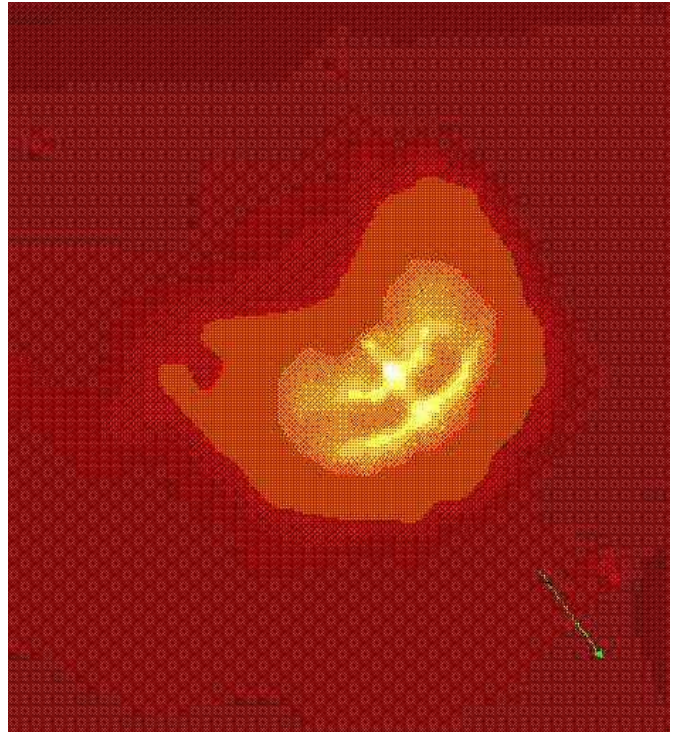,width=8.8cm,clip=} }
\caption{
Smoothed \cha\ image of the Vela Pulsar. 
In this 3.5-arcmin-square image, the green arrow denotes the proper-motion direction. 
The image is courtesy of NASA/PSU/Pavlov et al.\ (2001a).  
\label{f:vela_pulsar}}
\end{figure}

Fig.~\ref{f:vela_pulsar} shows the beautiful image (Pavlov et al. 2001a, 2001b) of the Vela Pulsar and the complex morphology~--- similar to that of the Crab, and B1509-58 (see below) --- of its surrounding nebula. 
In that this symposium concentrates on the neutron stars themselves, I shall not discuss the fascinating morphology of the pulsar-wind driven nebula here.
As with the Crab Pulsar, the angular resolution of \cha\ is essential to probing the detailed spectral properties of the pulsar and to isolate these from those of the nebula. 
Pavlov et al.\ performed observations using the Low-Energy Transmission Grating with the High-Resolution Camera (LETGS) and using the Advanced CCD Imaging Spectrometer (ACIS) back-illuminated CCDs, in order to characterize the spectral properties. 
Separate fits to LETGS and ACIS-S3 data determined the spectral parameters. 
However, the data were insufficient to discriminate fully between various spectral models~--- blackbody plus power law versus a hydrogen atmosphere plus power law. 
Nevertheless, an important result was the evidence for a harder spectral component~--- a thermal interpretation leading to "an implausibly small size" of 10 m (stretching even a quark-star interpretation!!)~--- thus indicating spectral similarities with middle-aged ($10^5 - 10^6$-y) X-ray pulsars.

\subsection{B1509-58}

\begin{figure}
\centerline{\psfig{file=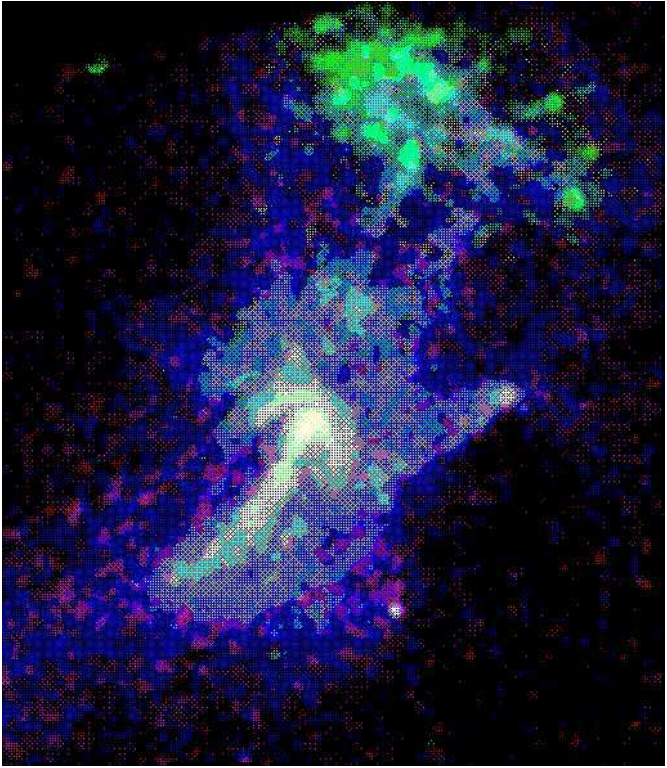,width=8.8cm,clip=} }
\caption{
\cha\ ACIS image of pulsar B1509-58 in SNR G230.4-1.2.
The colors in this $10\!\times\!14$-arcmin image indicate relative intensity. 
The image is courtesy NASA/MIT/Gaensler et al.  
\label{f:B1509-58}}
\end{figure}

Fig.~\ref{f:B1509-58} shows the young pulsar B1509-58, within SNR~G230.4-1.2 in the constellation Circinus.
This is another example of a pulsar-driven (plerionic) synchrotron nebula, the complex structure of which \cha\ resolves dramatically. 

\subsection{G292.0+1.8}

\begin{figure}
\centerline{\psfig{file=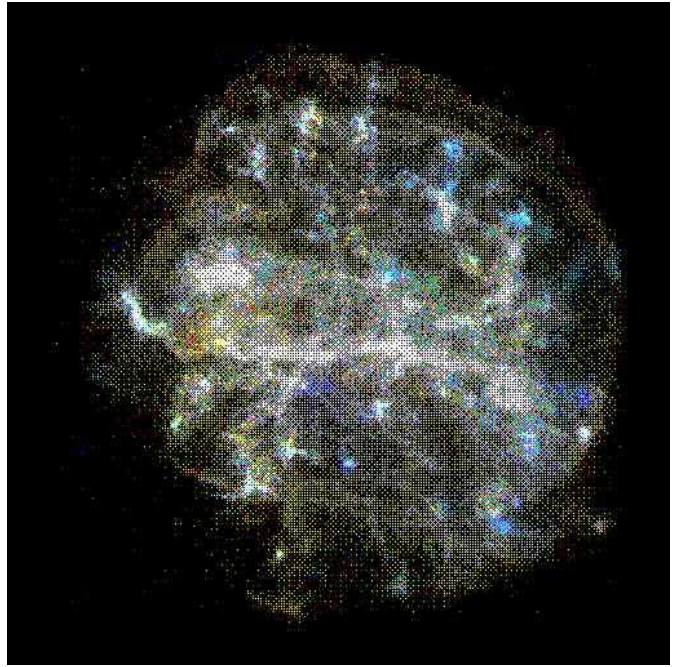,width=8.8cm,clip=} }
\caption{
\cha\ ACIS-S image of the oxygen-rich SNR G292.0+1.8, with the pulsar lying southeast of the center.
In this 9-arcmin-square image, the colors indicate relative intensity.  
The image is courtesy NASA/CXC/Rutgers/Hughes et al.  
\label{f:G292.0+1.8}}
\end{figure}

Along with Cas~A and Puppis, SNR G292.0+1.8, with a 1700-y estimated age (Murdin \& Clark 1979; Braun et al.\ 1983) is one of three known oxygen-rich supernovae in the Galaxy. 
Hughes et al.\ (2001) performed a moderately long ACIS--S3 observation (Fig.~\ref{f:G292.0+1.8}), detecting a bright, spectrally hard, point source within an apparently extended region.
This suggested the presence of a pulsar and its pulsar-wind nebula. 
Indeed, subsequent radio observations (Camillo et al.\ 2002) detected a 135-ms pulsar, noting that "X rays provide an important complement to the radio band, the traditional hunting ground of pulsar studies."
Camillo et al. (2002) also report that follow-up X-ray observations (Hughes et al.\ 2002, in preparation) confirm the presence of X-ray pulses with a period consistent with the radio observations. 
The X-ray spectroscopy is consistent with a simple power law, although, as with Vela, the model fit is not unique. 
It is interesting to note that the spin-down age is 2900 years, in contrast to the 1600--1700 years inferred from the supernova remnant.

\subsection{G11.2-0.3}

\begin{figure}
\centerline{\psfig{file=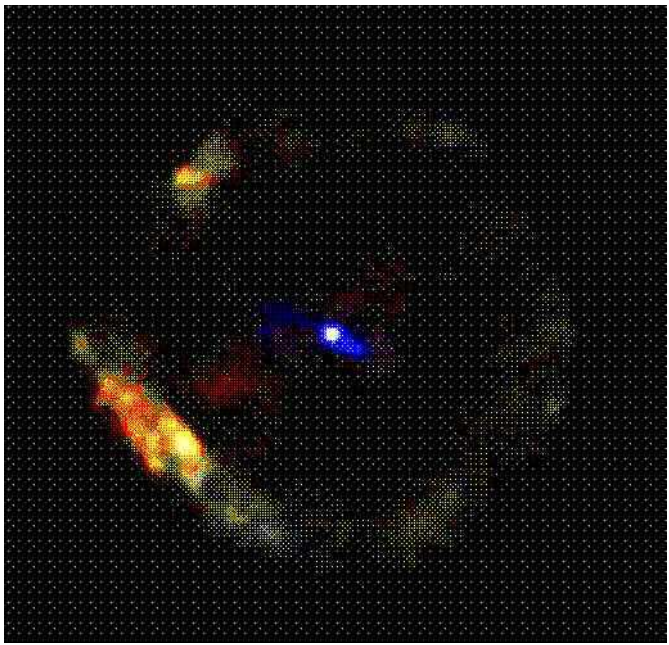,width=8.8cm,clip=}}
\caption{
\cha\ ACIS image of SNR G11.2-0.3, with its central pulsar and faint pulsar-wind inner nebula.
In this $5.8\!\times\!5.4$-arcmin image, the colors denote 0.6--1.65-keV (red), 1.65--2.25-keV (green), and 2.25--7.5-keV (blue) energy bands. 
The image is courtesy NASA/McGill/Kaspi et al. (2001) 
\label{f:G11.2-0.3}}
\end{figure}

Kaspi et al.\ (2001) used \cha\ to obtain the precise location of the 65-ms pulsar at the center of G11.2-0.3. 
The X-ray image (Fig.~\ref{f:G11.2-0.3}) reveals the presence of a pulsar-wind nebula around the pulsar, which lies within 8 arcsec of the center of the SNR shell, consistent with the historical supernova of 386. 
Because the spin-down age exceeds the historical age (as with G292.0+1.8) by a factor of 12, Kaspi et al.\ (2001) argue against the assumption that all pulsars are born with a very short period, as was the Crab Pulsar.
Invoking a larger birth period (closer to 62 ms) obviates the need for non-conventional spin-down.  

\subsection{IC443}

\begin{figure}
\centerline{\psfig{file=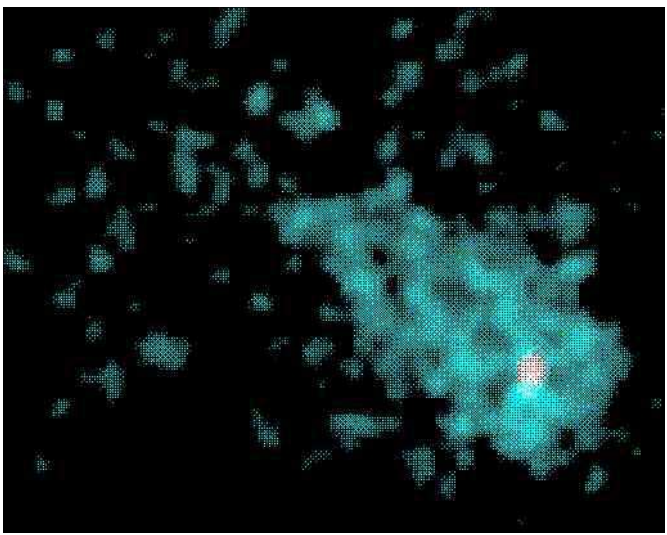,width=8.8cm,clip=} }
\caption{
\cha\ ACIS image of SNR IC443.
In this $1.0\!\times\!0.8$-arcmin image, the colors denote intensity. 
The image is courtesy NASA/NCSSM/Olbert et al.  
\label{f:ic443}}
\end{figure}

A sociologically, as well as scientifically, interesting \cha\ result was the detection of a point source in SNR IC443 (Fig.~\ref{f:ic443}) by high-school students from the (Durham) North Carolina School of Mathematics and Science. 
Their efforts led to winning the team award of the 2000 Siemens--Westinghouse Science and Technology Competition. 
(One of these students~--- Mr. Olbert, now at Columbia University~--- is presenting a poster paper at this symposium.)
The limited spectral data from this short (10-ks) observation are consistent, but not uniquely, with a 0.7-keV blackbody. 
(See, however, Bocchino \& Bykov 2001 for an XMM-{\sl Newton} viewpoint.)

\subsection{Others}

The above examples indicate the significance and utility of \cha\ observations of compact objects in SNRs. 
See also Kaspi et al.\ (2002) for a discussion of the detection of PSR B1757-24 and Pavlov et al.\ (2001b) for a discussion of the compact object in the RX J0852.0-4622 SNR.
I apologize for failing to reference other papers relevant to this topic. 

\section{Transients in Quiescence}

X-ray transients spend a considerable fraction of their time in quiescence.
Data from \cha\ observations have been used to test models of the emission mechanism in the quiescent state. 
We discuss here \cha\ observations of Cen X-4, Aquila X-1, and KS~1731-260. 

\subsection{Centaurus X-4 and Aquila X-1}

Rutledge et al.\ (2001a, 2001b) have analyzed \cha\ observations of Cen X-4 and Aquila X-1 in quiescence. 
Their analysis~--- discussed in terms of the detailed model by Brown, Bildsten, \& Rutledge (1998)~--- emphasizes that a blackbody spectrum is not an appropriate description of a weakly magnetic transient in quiescence. 
For Cen X-4, the authors find that no single-component spectral model produces a statistically acceptable fit.
While there is no unique two-component model, the authors use astrophysical arguments to interpret the data in terms of thermal plus power-law emission, deducing a radius for the neutron star.

For Aquila X-1, the spectral situation is both simpler and more complicated.
Although the data are consistent with thermal emission from a pure hydrogen atmosphere on the neutron star, models that include an additional power-law component (indicated by non-\cha\ observations) give equally acceptable fits. As with Cen X-4, the authors invoke astrophysical arguments to down-select spectral models. 

\subsection{KS 1731-260}

KS~1731-260 differs from the above-mentioned two X-ray transients, in the duration of the transient activity. 
The observation of type-I X-ray bursts from KS~1731-260 substantiate it's neutron-star identification. 
The principal difference between this source and Cen X-4 and Aquila X-1 is that its outburst phase is apparently longer~--- perhaps more than 12 years. 
Wijnands et al.\ (2001), who performed a 20-ks \cha\ ACIS-S observation during quiescence, summarize these and other characteristics.
The spectral fits do not distinguish statistically between a blackbody (kT = 0.3~keV) and a pure, very soft power-law model (photon index $> 4$). 
Their discussion, based on the blackbody fit and other assumptions, concludes that either the star has spent much of its life in quiescence for it to cool, or that neutrino cooling has been enhanced compared to Aquila~X-1. 

\section{Isolated Neutron Star RX J1856.5-3754}

Burwitz et al.\ (2001) obtained a 56-ks LETGS observation of the isolated neutron star RX J1856.5-3754. 
They found that none of the possibly relevant neutron-star atmosphere models, at least those neglecting magnetic fields, yield as good a spectral fit as a simple blackbody.
Furthermore, they found no evidence for electron cyclotron resonance lines.
Subsequent to these observations, \cha\ Director's Discretionary time increased the exposure to over 500 ks, with the data being immediately public. 
Early analysis of these data (Walter (2001); Burwitz, these proceedings) confirm the original results and set more stringent limits on the presence of any line features.
I am convinced that we have not begun to appreciate the implications of this observation. 
(Subsequent to the symposium, Drake et al.\ (2002) made the bold suggestion that these data may be interpreted as evidence for a quark star. 
However, see Pons et al.\ (2002) for an opposite view.)

\section{Conclusions}

High-spatial-resolution spectrometric images~--- such as those obtained by \cha~--- are necessary to enhance our understanding of neutron stars in SNR and of pulsar-wind nebulae. 
Higher-throughput but lower-angular-resolution X-ray missions~--- such as XMM-{\sl Newton} and the planned Constellation X~--- will benefit from the complementary capabilities of \cha's high angular resolution.

I am somewhat concerned that much of the current data do not adequately distinguish between spectral models. 
Certainly, one factor is simply the limited statistics of the earliest and typically short observations. 
As mentioned above, using \cha\ data to guide analysis of the far less photon-starved XMM-{\sl Newton} data will help, if statistical limitations underlie these ambiguities. 
As high-throughput missions drive down the statistical errors, it will be necessary to reduce uncertainties in the instrumental response and to develop even more detailed and accurate astrophysical models.
These are areas that obviously need more work.  
Nevertheless, these early \cha\ results are clearly stimulating and exciting. 
As we should expect, the wealth of \cha\ and XMM-{\sl Newton} data is challenging current views of pulsar and pulsar-wind-nebula emission mechanisms.

\vskip 0.4cm

\begin{acknowledgements}

I gratefully acknowledge support by NASA, by the Heraeus Foundation, and by the Max-Planck-Institut f\"ur Extraterrestriche Physik in Garching. In addition I thankfully acknowledge the comments and clarifications to the manuscript made by my colleague Steve O'Dell whose abilities to express his thoughts is writing are outstanding. 
\end{acknowledgements}

%\end{document}

\clearpage

\end{document}